\documentclass[12pt]{article}
\usepackage{graphicx}
\usepackage{atlasunit}
\newcommand\pubnumber{ATL-PHYS-PROC-2016-178}
\newcommand\pubdate{\today}

\newcommand{\figHeight}{2.2in}

\newcommand{\mate}[1]{\ensuremath{#1}}
\newcommand{\numUnit}[2]{\mate{#1 #2}}
\newcommand{\abs}[1]{\mate{|#1|}}
\newcommand{\ZWindow}[1]{\abs{#1-\mZ}}
\newcommand{\numErr}[2]{\mate{#1 \pm #2}}
\newcommand{\numErrUnit}[3]{\numErr{#1}{#2 #3}}
\newcommand{\Fig}[1]{Figure \ref{#1}}
\newcommand{\minPtCut}[1]{\mate{\pT > \numUnit{#1}{\GeV}}}
\newcommand{\sqrtS}[1]{\mate{\sqrt{s} = #1 \TeV}}

\newcommand{\invfb}{\mate{\ \ifb}}
\newcommand{\pTlb}{\mate{p_{\mathrm{T}, \ell b}}}

\newcommand{\pT}{\mate{p_\mathrm{T}}}
\newcommand{\pTMiss}{\mate{\vec{p}_\mathrm{T}^\mathrm{miss}}}
\newcommand{\ETMiss}{\mate{E_\mathrm{T}^\mathrm{miss}}}
\newcommand{\mass}{\mate{m}}
\newcommand{\mtop}{\mate{\mass_\mathrm{top}}}
\newcommand{\mlb}{\mate{\mass_{\ell b}}}
\newcommand{\mll}{\mate{\mass_{\ell\ell}}}
\newcommand{\mZ}{\mate{\mass_\mathrm{Z}}}

\newcommand{\Rj}{\mate{R_{3/2}}}
\newcommand{\ChiSquared}{\mate{\chi^2}}
\newcommand{\MWMC}{\mate{\mass_{W}^\mathrm{MC}}}

\newcommand{\ttbar}{\mate{t\bar{t}}}
\newcommand{\MCSims}{Monte-Carlo simulations}
\newcommand{\bjet}{\mate{b} jet}
\newcommand{\leptonbjet}{lepton$-b$-jet}

\def\institute{Physikalisches Institut\\
Universit\"at Bonn, D-53115 Bonn, GERMANY}
\newcommand{\awkNotice}{The research leading to these results has received funding 
from the European Research Council under the European Union's Seventh Framework 
Programme ERC Grant Agreement n. 617185.}
\def\support{\footnote{\awkNotice}}

\def\Title#1{\begin{center} {\Large #1 } \end{center}}
\def\Author#1{\begin{center}{ \sc #1} \end{center}}
\def\Address#1{\begin{center}{ \it #1} \end{center}}

\newcommand\pubblock{\rightline{\begin{tabular}{l} \pubnumber\\
         \pubdate  \end{tabular}}}
\newenvironment{Abstract}{\begin{quotation}  }{\end{quotation}}
\newenvironment{Presented}{\begin{quotation} \begin{center} 
             PRESENTED AT\end{center}\bigskip 
      \begin{center}\begin{large}}{\end{large}\end{center} \end{quotation}}
\def\Acknowledgements{\bigskip  \bigskip \begin{center} \begin{large}
             \bf ACKNOWLEDGEMENTS \end{large}\end{center}}
\def\beq{\begin{equation}}
\def\eeq#1{\label{#1}\end{equation}}
\def\eeqn{\end{equation}}

\def\beqa{\begin{eqnarray}}
\def\eeqa#1{\label{#1}\end{eqnarray}}
\def\eeqan{\end{eqnarray}}

\let\bar=\overbar

\def\Dslash{\not{\hbox{\kern-4pt $D$}}}
\def\dslash{\not{\hbox{\kern-2pt $\del$}}}

\def\msb{{\bar{\ssstyle M \kern -1pt S}}}

\begin{document}
\begin{titlepage}
\pubblock

\vfill
\Title{New results on top-quark mass, including \\new methods, in ATLAS}
\vfill
\Author{ Kaven Yau Wong, On behalf of the ATLAS Collaboration\support}
\Address{\institute}
\vfill
\begin{Abstract}
Recent results on top-quark mass measurements with the ATLAS detector 
using proton-proton collisions at the Large Hadron Collider are presented.
These results correspond to the measurements in the \ttbar\ all-hadronic and dilepton channels
at \sqrtS{8}\ collisions and an integrated luminosity of \numUnit{20}{\invfb}.
\end{Abstract}
\vfill
\begin{Presented}
$9^{th}$ International Workshop on Top Quark Physics\\
Olomouc, Czech Republic,  September 19--23, 2016
\end{Presented}
\vfill
\end{titlepage}
\def\thefootnote{\fnsymbol{footnote}}
\setcounter{footnote}{0}

\section{Introduction}
The top quark is the heaviest elementary particle in the Standard Model (SM) and its
mass is a fundamental parameter that needs to be determined experimentally.
The top-quark mass has an important effect in electroweak radiative corrections and
a precise measurement is relevant, in particular, for theories of physics beyond the SM.
Furthermore, its value is an important test for the consistency of the SM.

The ATLAS detector \cite{ATLAS} is a general purpose detector 
located at the Large Hadron Collider (LHC). %\cite{LHC}.
Two top-quark mass measurements were performed in the last year using data from the ATLAS detector at
a center-of-mass energy of \sqrtS{8}:

\begin{itemize}
  \item the top-quark mass measurement in the dilepton channel \cite{massDilepton} and
	\item the top-quark mass measurement in the all-hadronic channel \cite{massAllHad}.
\end{itemize}

Both measurements use the template method to extract the top-quark mass, 
where the templates are derived using \MCSims.

\section{Top-quark measurement in the dilepton channel}
This analysis \cite{massDilepton} uses the full \sqrtS{8}\ ATLAS data, 
which gives an integrated luminosity of \numUnit{20}{\invfb}.

Although the \ttbar\ dilepton channel has a small branching ratio (6\%), 
this is compensated by an extremely high purity, of the order of 99\%.
The main disadvantage is the presence of two neutrinos which 
makes the full reconstruction of the event difficult, 
since the missing transverse momentum (\pTMiss) and the missing transverse energy (\ETMiss)
can only be associated to the combined effects of two particles.
The use of the template method circumvents this problem, 
since it exploits the expected distribution of the \mlb\ variable to extract the top-quark mass.

The \mlb\ variable is defined as the invariant mass of the two \leptonbjet\ pairs.
Since the correct pairing between each lepton and their corresponding \bjet\ is not known a priori, 
both combinations are computed and the combination giving the smallest value of \mlb\ is taken as the correct pairing.

In order to select signal events and reject background, the events are required to
have exactly two reconstructed leptons with opposite-sign charges,
at least two reconstructed jets and
at least one reconstructed jet must be $b$-tagged (tagger efficiency: 70\%).
All the reconstructed leptons and jets must have a transverse momentum (\pT) larger than \numUnit{25}{\GeV}.
For the $ee$ and $\mu\mu$ channels, it is also required that \mate{\ETMiss>\numUnit{60}{\GeV}} and
the invariant mass of the two reconstructed leptons (\mll) must satisfy 
\mate{\mll>\numUnit{15}{\GeV}} and \mate{\ZWindow{\mll} > \numUnit{10}{\GeV}}, where \mZ\ is the mass of the $Z$ boson.
For the $e\mu$ channel, the scalar sum of the \pT\ of all reconstructed jets and leptons must be larger than \numUnit{130}{\GeV}.

\begin{figure}[htb]
	\centering
	\includegraphics[height=\figHeight]{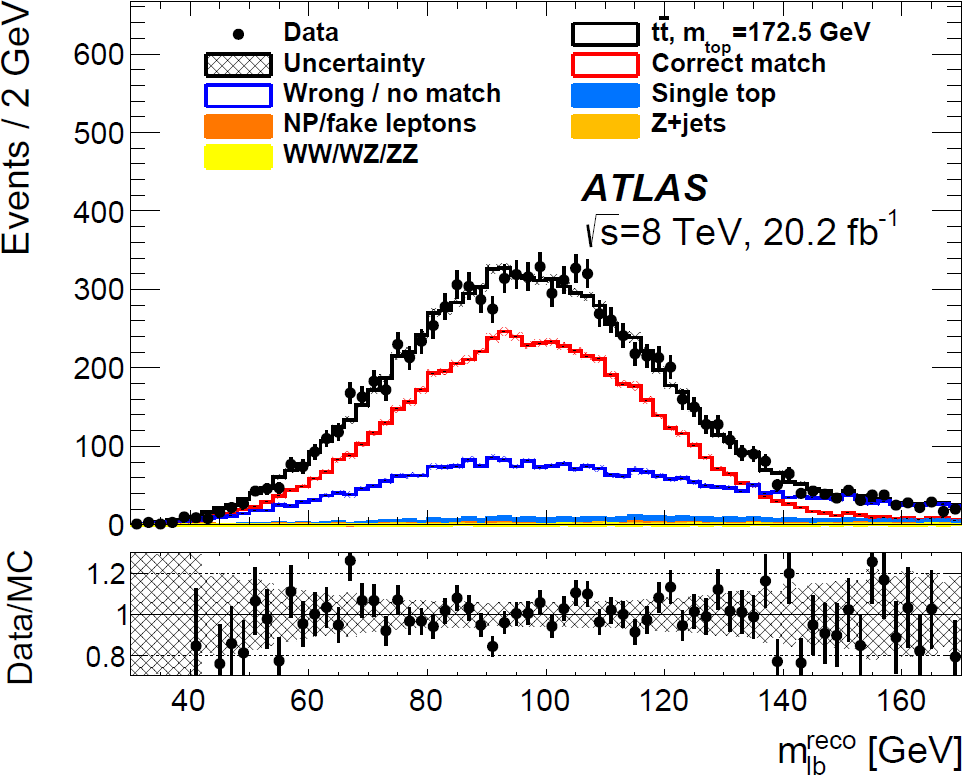}
	\includegraphics[height=\figHeight]{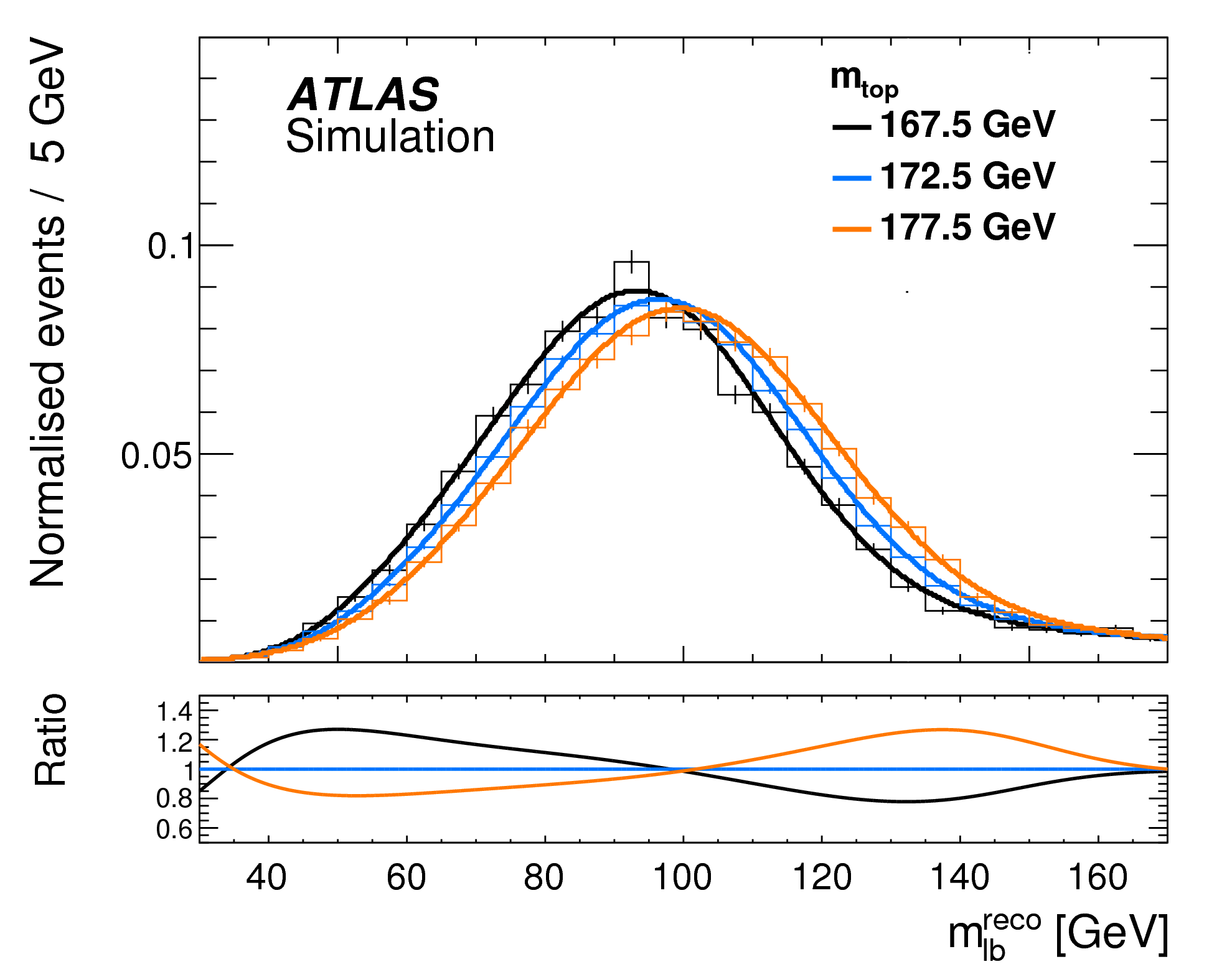}
	\caption{Left: comparison between data and simulation of the \mlb\ distribution after applying all cuts.
	Right: template used to measure the top-quark mass using the \mlb\ variable. 
	The histograms show the distribution generated by the \MCSims\ while the solid lines
	show the respective template distribution \cite{massDilepton}.}
	\label{figDil}
\end{figure}

In order to increase the pairing efficiency of the \leptonbjet\ pairs, events are also required to satisfy
\mate{\numUnit{30}{\GeV}<\mlb<\numUnit{170}{\GeV}} and the average transverse momentum of the two \leptonbjet\ pairs (\pTlb)
must be larger than \numUnit{120}{\GeV}. The distribution of \mlb\ after applying all the cuts is shown in \Fig{figDil} (left).

After applying the final selection, \numErr{10100}{770} events are expected,
of which \numErr{10030}{770} are predicted to be signal events.
The expected matching efficiency for the \leptonbjet\ pairing is \mate{(\numErr{95.3}{0.4})}\%.
Applying this selection to the data, 9426 events are found.

In order to measure the top-quark mass, the template shown in \Fig{figDil} (right) is used. 
This template is created by modelling the signal as the sum of a Gaussian and a Landau distribution,
while the background is modelled with a Landau distribution. 
The final template depends only on the top-quark mass.

Fitting this template to the data, a measurement of:
\[
	\mtop=\numUnit{172.99\ \pm\ 0.41\mathrm{(stat.)}\ \pm\ 0.74\mathrm{(syst.)}}{\GeV}
\]
is obtained, where the systematic uncertainty is dominated by the jet energy scale (\numUnit{0.54}{\GeV}).

This result is combined with the ATLAS top-quark mass measurements in the single-lepton and dilepton channels
performed at \sqrtS{7} \cite{ATLAS7TeV} using the Best Linear Unbiased Estimate method \cite{BLUE}.
The combined measurement gives a combined top-quark mass value of:
\[
	\mtop=\numUnit{172.84\ \pm\ 0.34\mathrm{(stat.)}\ \pm\ 0.61\mathrm{(syst.)}}{\GeV}.
\]

\section{Top-quark mass measurement in the all-hadronic channel}
This mass measurement \cite{massAllHad} also uses the full \sqrtS{8}\ ATLAS data, 
which gives an integrated luminosity of \numUnit{20}{\invfb}.

In the all-hadronic channel, both $W$ bosons decay hadronically, giving a signature of four light jets and two \bjet s.
Unlike the dilepton channel, the all-hadronic channel has the advantage of having the largest branching ratio of all channels,
no neutrinos and, hence, the ability to perform a full kinematic reconstruction of the event.
The main disadvantage of the all-hadronic channel is the large amount of jets, 
which poses a complex combinatorics problem to properly identify and reconstruct events. 
Furthermore, the kinematic reconstruction of an event depends heavily on the jet energy scale.
Finally, the multijet background is significant.

The \Rj\ variable is used for the template method. 
It is defined, in a top quark hadronic decay, as the ratio between the invariant mass 
of the three jets (one \bjet\ and two coming from the decay of the $W$ boson) and the
invariant mass of the two jets that are the product of the $W$-boson decay.
Since there are two top-quark decays per \ttbar\ event, two \Rj\ values can be computed per event.
The correlation between the two values of \Rj\ per event is 0.59 and 
such correlation is considered in the estimation of the uncertainties.

The events are required to have no reconstructed leptons,
at least six reconstructed jets with \minPtCut{25}, 
of which at least five reconstructed jets must have \minPtCut{60}
and at least two of the six leading-\pT\ jets must be $b$-tagged (tagger efficiency: 57\%).
It is also required that \mate{\ETMiss<\numUnit{60}{\GeV}} and that 
the azimuthal separation between the two \bjet s with the highest $b$-tagging weights must be larger than 1.5.
Furthermore, after performing the \ttbar\ kinematic fit explained in the next paragraph, 
the average azimuthal separation between the corresponding \bjet s and $W$ bosons of both decay chains must be smaller than 2
and the smallest value of \ChiSquared\ must be less than 11. The final selection gives an expected purity of 34\%.

In order to fully reconstruct the \ttbar\ event, a kinematic fit is performed by minimizing the value of:
\[
	\ChiSquared = \frac{\left(\mass_{b_1 j_1 j_2} - \mass_{b_2 j_3 j_4}\right)^2}{\sigma^2_{\Delta m_{bjj}}} +
	\frac{\left(\mass_{j_1 j_2} - \MWMC \right)^2}{\sigma^2_{\MWMC}}+
	\frac{\left(\mass_{j_3 j_4} - \MWMC \right)^2}{\sigma^2_{\MWMC}},
\]
where \mate{b_1} is the \bjet\ originating from the top quark decay, 
\mate{b_2} is the \bjet\ originating from the antitop quark decay, 
\mate{j_1} and \mate{j_2} are the jets originating from the \mate{W^{+}} decay, while
\mate{j_3} and \mate{j_4} are the jets originating from the \mate{W^{-}} decay.
The values of \mate{\MWMC = \numErrUnit{81.18}{0.04\mathrm{(stat.)}}{\GeV}}, 
\mate{\sigma_{\Delta m_{bjj}} = \numErrUnit{21.60}{0.16\mathrm{(stat.)}}{\GeV}} and 
\mate{\sigma_{\MWMC} = \numErrUnit{7.89}{0.05\mathrm{(stat.)}}{\GeV}} are determined from 
\MCSims\ using the correct combination of jets, which is obtained from the event generator.

During the \ttbar\ event reconstruction, all the possible jet combinations are tried, 
and the combination giving the smallest value of \ChiSquared\ is considered 
the correct jet combination of the event.

The QCD multijet background is the largest background contribution and it is estimated using data-driven methods.
Its uncertainty is expected to have an impact of \numUnit{0.16}{\GeV} in the final top-quark mass measurement.

\begin{figure}[htb]
	\centering
	\includegraphics[height=\figHeight]{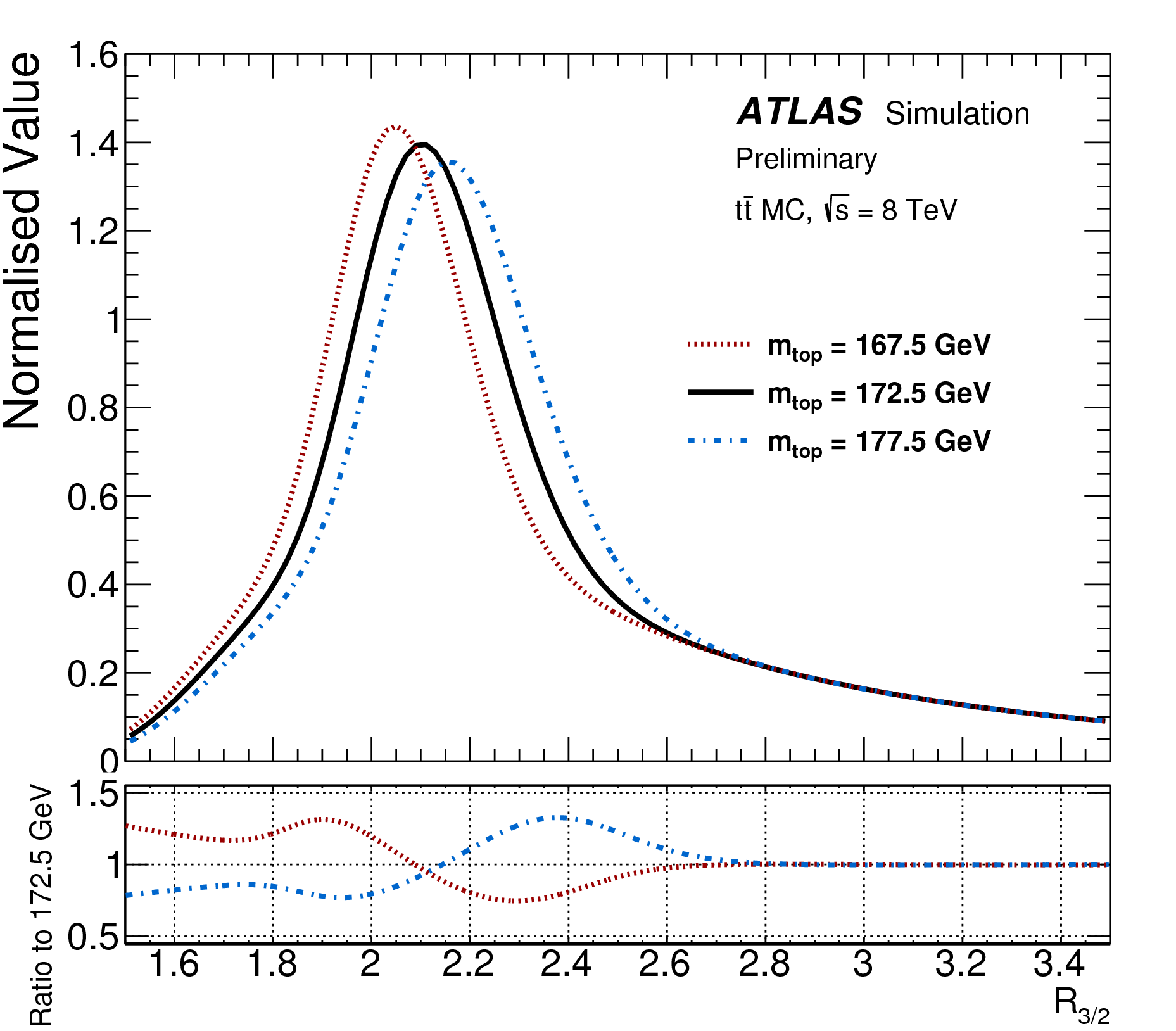}
	\includegraphics[height=\figHeight]{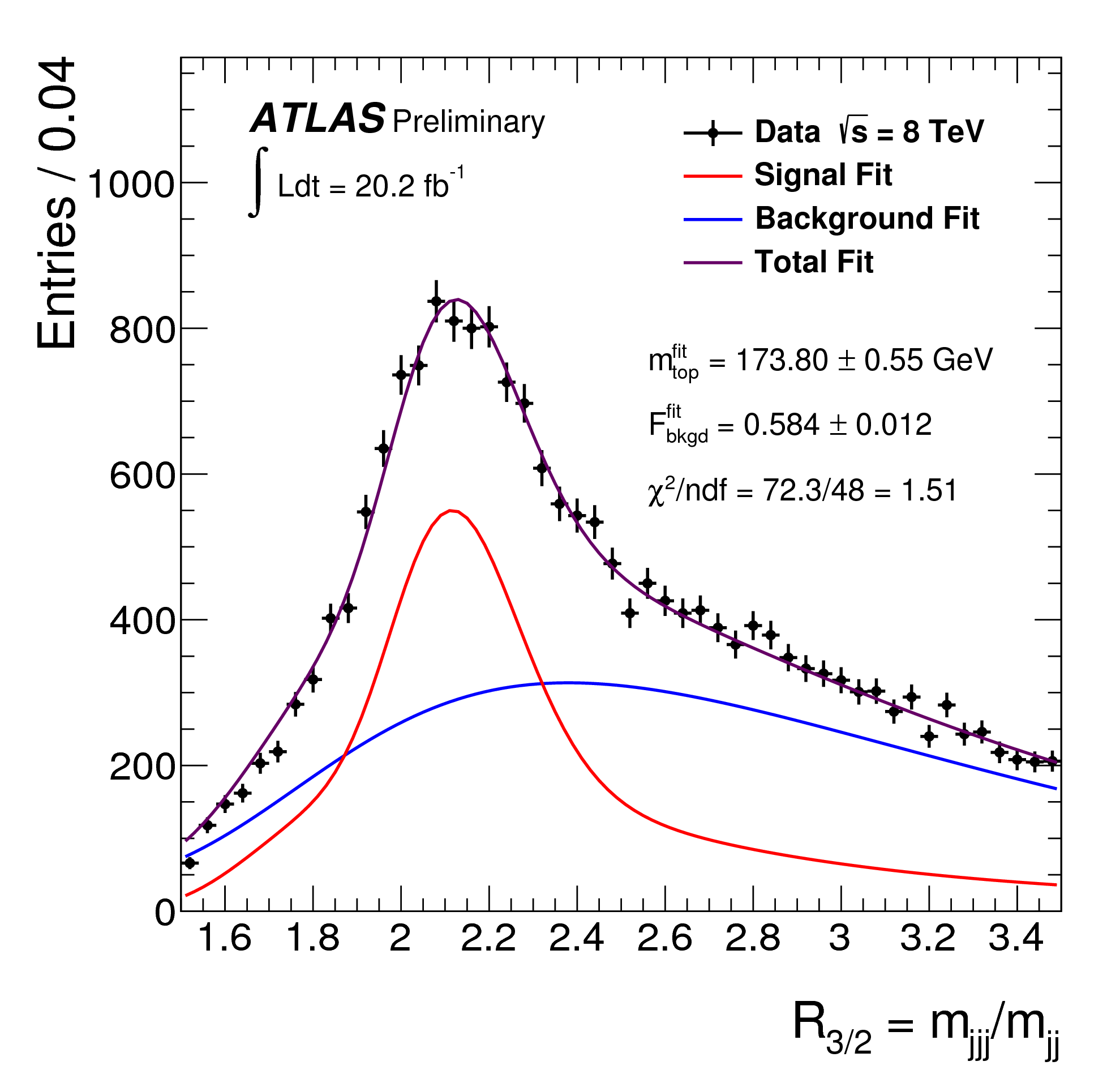}
	\caption{Left: template used to measure the top-quark mass using the \Rj\ variable. 
	Right: result of the template fit to the distribution of the \Rj\ variable in data \cite{massAllHad}.}
	\label{figHad}
\end{figure}

In order to measure the top-quark mass, the template shown in \Fig{figHad} (left) is used. 
This template is created by modelling the signal distribution with a Novosibirsk distribution,
while the background distribution is modelled with a Landau distribution. 
The final template depends on the top-quark mass and the background fraction parameter (\mate{F_\mathrm{bkgd}}).

Fitting the template to the data, as shown in \Fig{figHad} (right), a top-quark mass of
\[
	\mtop=\numUnit{173.80\ \pm\ 0.55\mathrm{(stat.)}\ \pm\ 1.01\mathrm{(syst.)}}{\GeV}
\]
is measured, where the systematic uncertainty is dominated by the 
hadronization modelling (\numUnit{0.64}{\GeV}) and the jet energy scale (\numUnit{0.60}{\GeV}).

\Acknowledgements
\awkNotice


\begin{thebibliography}{99}
\bibitem{ATLAS} 
ATLAS Collaboration, 
The ATLAS Experiment at the CERN Large Hadron Collider, 
JINST 3 (2008) S08003.

%\bibitem{LHC}
%L. Evans and P. Bryant (editors),
%LHC Machine,
%JINST 3 (08) (2008) S08001.

\bibitem{massDilepton}
ATLAS Collaboration,
Measurement of the top quark mass in the \ttbar \mate{\rightarrow} dilepton channel from \sqrtS{8}\ ATLAS data,
Phys. Lett. B761 (2016) 350-371.

\bibitem{massAllHad}
ATLAS Collaboration,
Measurement of the top quark mass in the all-hadronic \ttbar\ decay channel at \sqrtS{8}\ with the ATLAS detector,
ATLAS-CONF-2016-064, http://cds.cern.ch/record/2206204.

\bibitem{ATLAS7TeV}
ATLAS Collaboration,
Measurement of the top quark mass in the \mate{\ttbar \rightarrow \mathrm{lepton+jets}} and \mate{\ttbar \rightarrow \mathrm{dilepton}} channels using \sqrtS{7}\ ATLAS data,
Eur. Phys. J. C 75 (2015) 330.

\bibitem{BLUE}
L. Lyons et al., 
How to combine correlated estimates of a single physical quantity, 
Nucl. Instr. Meth. A 270 (1988) 110.

\end{thebibliography}
\end{document}